\documentclass[aps,prb,floats,twocolumn,showpacs,amsmath,amsfonts,amssymb]{revtex4}
\usepackage{graphicx}
\usepackage{subfigure}
\usepackage{dcolumn}% Align table columns on decimal point
\usepackage{bm}% bold math
\begin{document}
%%%%%%%%%%%%%
\title{Negative differential conductance in quantum dots in theory and
  experiment}
%%%%%%%%%%%%%
\author{M. C. Rogge$^{1}$, F. Cavaliere$^{2,3}$, M.  Sassetti$^{2}$,
R. J. Haug$^{1}$, and B. Kramer$^{3}$ \vspace{1mm}}
\affiliation{$^{1}$Institut f\"ur Festk\"orperphysik,
Universit\"at Hannover, Appelstra\ss{}e 2, 30167 Hannover, Germany\\
  $^{2}$ I. Institut f\"ur Theoretische Physik, Universit\"at Hamburg,
  Jungiusstra\ss{}e 9, 20355 Hamburg, Germany\\
  $^{3}$INFM-LAMIA, Dipartimento di Fisica Universit\`{a}
  di Genova, Via Dodecaneso 33, 16146 Genova, Italy}
  \date{\today}
%%%%%%%%%%%%%%%%%%%%%%%%%%%%%%%%%%%%%%%%%%%%%%%%%%%%%%%%%%%%%%%%%%%%
\begin{abstract}
  Experimental results for sequential transport through a lateral quantum dot
  in the regime of spin blockade induced by spin dependent tunneling are
  compared with theoretical results obtained by solving a master equation for
  independent electrons. Orbital and spin effects in electron tunneling in the
  presence of a perpendicular magnetic field are identified and discussed in
  terms of the Fock-Darwin spectrum with spin. In the nonlinear regime, a
  regular pattern of negative differential conductances is observed.
  Electrical asymmetries in tunnel rates and capacitances must be introduced
  in order to account for the experimental findings. Fast relaxation of the
  excited states in the quantum dot have to be assumed, in order to explain
  the absence of certain structures in the transport spectra.
\end{abstract}
 \maketitle
%%%%%%%%%%%%%%%%%%%%%%%%%%%%%%%%%%%%%%%%%%%%%%%%%%%%%%%%%%%%%%%%%%%%
%%%
%%%
%%% INTRODUCTION
%%%
%%%
%%%%%%%%%%%%%%%%%%%%%%%%%%%%%%%%%%%%%%%%%%%%%%%%%%%%%%%%%%%%%%%%%%%%
Since the discovery of the Coulomb blockade effect in semiconductors
\cite{Meirav-PRL-90} electronic transport properties of quantum dots have been
the subject of continuous interest. The interplay of Coulomb interaction
between quantized charges, the spectra of orbital states and the electron spin
causes a rich variety of effects in the transport spectra of lateral quantum
dots \cite{Kouwenhoven-97}. In an external magnetic field, signatures have
been found for transitions between different angular momentum states of the
single electron Fock-Darwin spectrum \cite{Fock-ZP-28,Darwin-MPCPS-30},
modified by interaction effects \cite{McEuen-PRB-92}. Effects of the electron
spin have been observed, including several transport blocking mechanisms. One
of these spin blockades was introduced by Ciorga et al.
\cite{Ciorga-PRB-00,Ciorga-PRL-02}. Due to spin polarized leads Coulomb peaks
show a spin dependent amplitude. This was observed in split gate quantum dots
and recently in devices made with local anodic oxidation \cite{Rogge-APL-04}.
In nonlinear measurements negative differential conductance was observed and
explained in terms of spin dependent tunneling\cite{Ciorga-APL-02}.

In the present paper, we report results of spin blockade
measurements of the linear and nonlinear sequential transport of a
2D quantum dot in a perpendicular magnetic field in comparison
with a theoretical model. In the considered parameter region, we
confirm results concerning angular momentum and spin blockade
effects obtained
earlier\cite{Ciorga-PRB-00,Ciorga-PRL-02,Ciorga-APL-02}. In
particular, we observe the characteristic zig-zag patterns of the
linear conductance peaks when varying the magnetic field. The
positions and the strengths of the linear conductance peaks show a
regular alternating behavior as a function of gate voltage and
magnetic field. In the nonlinear transport regime a complicated
substructure involving certain transitions to excited states is
revealed. A striking bi-modal pattern of positive and negative
differential conductance (PDC and NDC) is observed below the
$(N+1)\to N$ transition line which is the extension of the linear
zig-zag pattern. This appears to be associated with a strong
depletion of the conductance peaks corresponding to transitions
$N\to (N+1)$. To study the conditions for obtaining the NDC in the
nonlinear transport induced by the spin blockade we compare the
experimental findings with theoretical results for the nonlinear
current-voltage characteristics that were obtained by solving the
rate equation for the Fock-Darwin spectrum with suitably defined
spin dependent tunnel rates. In addition to the linear conductance
we investigated in detail the conditions for the spin induced NDC
to appear in the nonlinear differential conductance. It is shown
that assuming spin dependent tunneling only is {\em not}
sufficient to understand the data. Our main result is that it is
necessary to assume a rather complex interplay of asymmetries, and
the {\em relaxation of certain excited states} to be much faster
than the transition rates between leads and core states of the
quantum dot, in order to account for the regularity in the
experimental conductance traces in a large parameter region.
\begin{figure}
\includegraphics[scale=1]{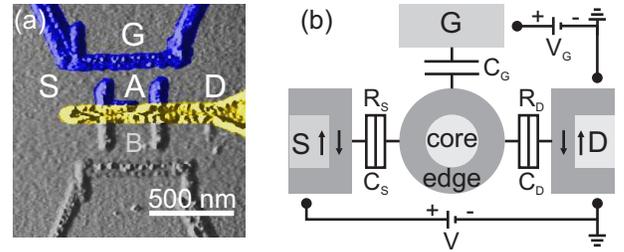}
\caption{(a) AFM picture of the gate
  structures used to define electrostatically a lateral quantum dot in the
  inversion layer of a AlGaAs/GaAs heterostructure. (b) electrostatic model
  for the device in a perpendicular magnetic field. The dot with edge and core
  is coupled via tunnel resistances R and capacitances C to source S and drain
  D that feature spin split edge channels. The in-plane gate G is coupled
  capacitively.} \label{fig1}
\end{figure}

The quantum dot has been fabricated in an inversion layer embedded
in a GaAs/AlGaAs heterostructure. The 2D electron system (2DES) is
located 34~nm below the surface. The sheet density is $n=4.3 \cdot
10^{15}$~m$^2$. A combination of Local Anodic Oxidation (LAO) with
an Atomic Force Microscope (AFM) and electron beam lithography
(ebeam) is used to define the quantum dot structure visible in
Fig.~\ref{fig1}a. Details of the sample processing can be found in
[10,11].
%\cite{Rogge-APL-03,Rogge-PE-04}
Quantum dot A is coupled to source (S) and drain (D) leads via
resistances $R_S$ and $R_D$ and capacitances $C_S$ and $C_D$
(Fig.~\ref{fig1}b). An in-plane gate G is coupled via $C_G$ and
can be biased to tune the tunneling rates and electron numbers of
the dot. Although the dot B is coupled capacitively to A the
measurements discussed here reflect the dynamics of dot A only.
They are performed in a regime with dot B only connected to the
source lead. Thus transport can occur only via dot A. The
influence of the second dot is discussed in [8].
%\cite{Rogge-APL-04}
With dot A in the Coulomb blockade regime the differential conductance is
recorded using standard lock in technique at a temperature of 50~mK.

\begin{figure}
\includegraphics[scale=1]{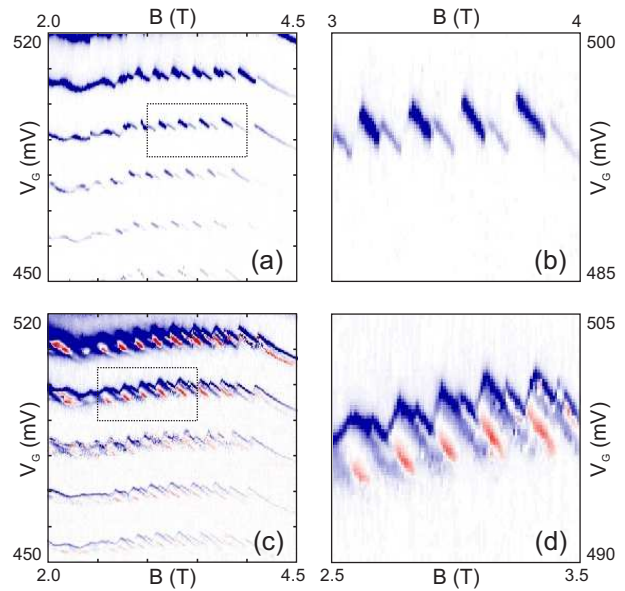}
\caption{(a) Linear conductance peaks of the quantum dot in the
  plane of gate voltage $V_G$ and magnetic field $B$ (dark: high
  positive differential conductance, bright: low conductance); (b) enlarged
  inset of (a). The Coulomb peaks develop a bimodal behavior in position and
  amplitude. (c) spectrum of the nonlinear conductance in the same parameter
  region at bias $V=0.5$\,mV  (red: NDC); (d)
  enlarged inset of (c). A regular pattern of PDC and NDC is
  observed.}\label{fig2}
\end{figure}

Figures \ref{fig2}a,b show typical results for the differential
conductance in the plane of $V_G$ and $B$ with $V=0$. The
sawtooth-like traces that are separated by the charging energy
show several characteristic features. Firstly, there is a clear
{\em bimodal behavior} of the peak position. An oscillation with a
huge increase follows one with a small increase. The same is true
for the intensity of the conductance due to spin dependent
transport. Secondly, the intensity of all traces with {\em
positive slopes is very weak}, almost not visible in this plot.
Thirdly, traces corresponding to successive Coulomb peaks are {\em
shifted relative to each other} by half a period. Large
intensities in one trace correspond to small and weak features in
the neighboring traces. In the nonlinear conductance
(Fig.~\ref{fig2}c,d), such features are also present.
Additionally, one notes that while the large and strong features
are associated with positive differential conductance in the
nonlinear region, the smaller and weaker features are continued
into the nonlinear regime with NDC. An important observation is
that only traces with a {\em negative} slope as a function of the
magnetic field are observed to yield NDC.

In order to understand these features, we consider a model for $N$
uncorrelated electrons in a 2D harmonically confined quantum dot.
The ground state energy for $N$ electrons in this system is $
H_{N}=E_C{N}^{2}/2-eNV_G{C_G}/{C_{\Sigma}}-eNV{C_S}/{C_{\Sigma}}+\sum_{j=1}^{N}\varepsilon_{j}(B)$.
The first term is the classical charge addition energy ($E_C$
charging energy) which accounts for the Coulomb repulsion.
$C_{\Sigma}=C_S+C_D+C_G\approx C_S+C_D$ is the total capacitance
of the dot. The lever arm $\alpha=C_G/C_{\Sigma}$ accounts for the
energy change induced by $V_G$. The last term is the kinetic
energy contributed by occupying the energetically lowest
one-electron states of the quantum dot with $N$ electrons. The
chemical potential for the transition $N-1\leftrightarrow N$ is
defined as
$\mu_{N}(B):=H_{N}-H_{N-1}=E_C(N-1/2)+E_{N}(B)-eV_G{C_G}/{C_{\Sigma}}-eV{C_S}/{C_{\Sigma}}$.
For the nonlinear regime also the lowest lying excited states with
energies $H_{N}^{*}=H_{N}+\varepsilon_{N+1}-\varepsilon_{N}$ must
be considered.

\begin{figure}
\includegraphics[scale=1]{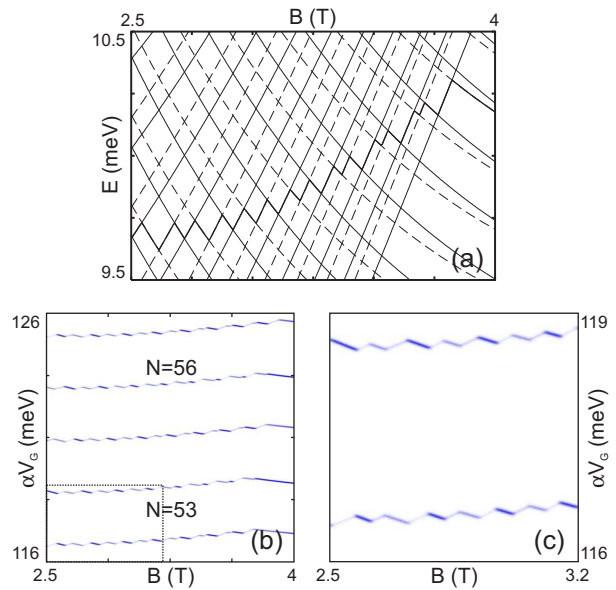}
\caption{(a) Fock-Darwin energy spectrum of a quantum dot as a
  function of the magnetic field $B$ in the $2\leq \nu\leq 4$ regime: states
  with spin $-1$ ($+1$) correspond to solid (dashed) lines; bold:
  $E_{57}(B)$. (b) Theoretically evaluated linear conductance traces as a
  function of $V_{\rm G}$ and $B$ for the transitions $N=52\to
  53$,\ldots,$N=56\to 57$ with $\gamma_{\uparrow}=0.5$ and $\gamma_c=0.2$ (see
  text). (c) Enlarged region of (b); parameters: $\hbar\omega_{0}=1.3\
  \textrm{meV}$, $g^*=0.44$, $T=70\ \textrm{mK}$.} \label{fig3}
\end{figure}

For harmonic confinement, $
\varepsilon_{j}(B)=E_{j}(B)+s_{j}E_z(B)$ with $E_z=g^{*}\mu_BB$.
The first term represents the $j$-th level of the Fock-Darwin (FD)
spectrum, while the second term describes the Zeeman splitting of
the FD spectrum with $g^{*}$ the electron $g$-factor, the Bohr
magneton $\mu_B$, and $s_{j}=\pm 1$ the $z$ component of the spin
of the $j$-th FD level (units $\hbar/2$). Here, we assume
$\varepsilon_{1}(B)\leq\varepsilon_{2}(B)\leq\ldots$. For small
$B$, the ground state is a spin singlet or triplet depending on
whether $N$ is even or odd, respectively. Increasing the magnetic
field, many level-crossings occur in the FD spectrum. This is
shown in Fig.~\ref{fig3}a where the FD spectrum and the energy
level $E_{57}(B)$ are shown for dot filling factor ($\nu$)
$2\leq\nu\leq 4$. For very large magnetic field,
$\omega_c=eB/m^*\gg \omega_{0}$ ($m^{*}$ electron effective mass,
$\omega_{0}$ characteristic frequency of harmonic confinement
potential), the Fock-Darwin levels merge into almost degenerate,
Zeeman split Landau bands.

The dot is connected to external noninteracting leads by means of tunneling
junctions with resistances $R_{\rm{S,D}}$ and capacitances $C_{\rm{S,D}}$
(Fig.~\ref{fig1}b). The source (S) terminal is kept at a potential $V$, while
the drain (D) is grounded. At low temperature, for $V>0$ electrons flow
essentially from the drain to the source lead. We consider the tunneling in
the sequential regime and introduce the tunneling rates $\Gamma_{i\to
  f}\propto\gamma_{i,f}\Gamma_{0}$ with $\Gamma_0$ in dimension of a tunneling
rate and $\gamma_{i,f}$ dimensionless parameters which depend on
the states involved in the tunneling process. In particular, we
have found to be useful to introduce the following parameters:
$\gamma_{S,D}$ tunneling via source or drain; $\gamma_{e,c}$
tunneling via edge or core ground state transitions;
$\gamma_{e,c}^*$ tunneling via edge or core excited states
transitions; $\gamma_{\downarrow,\uparrow}$ tunneling with spin
down or spin up.  For example, for tunneling with spin up via the
source to an excited state in the core we have
$\gamma_{i,f}=\gamma_{\uparrow}\cdot\gamma_S\cdot\gamma_c^*$
(throughout this work, $\gamma_e=1$, $\gamma_e^* =1$ and
$\gamma_{\downarrow}=1$).

We have calculated the current and the differential conductance by
solving a master equation. First, we address the linear transport
regime where only ground states $N$, $N+1$ are involved. The
transition $N\leftrightarrow N+1$ occurs if the resonance
condition $\mu_{N+1}(B)\approx 0$ is fulfilled. Therefore, the
peak position in the plane $(B,V_{\rm G})$ reflects the position
as a function of $B$ of the $N+1$-th energy level. By comparing
the linear conductance traces in Fig.~\ref{fig2}a and the number
of kinks in the trace of the chemical potential in
Fig.~\ref{fig3}a, one can conclude that the number of electrons is
in the range $N=53,\ldots,57$. In the $2\leq\nu\leq 4$ regime, the
dot states with an increasing (decreasing) $E_{j}(B)$ have a low
(high) angular momentum and are located in the core (edge) of the
dot (Fig.~\ref{fig1}b). This leads to a weaker (stronger)
overlapping of the electron wavefunction with the leads states. To
take into account this fact, tunneling rates involving a low
angular momentum state are characterized by the factor
$0<\gamma_c<1$. Moreover, because of the spatial separation of
spin-resolved edge states in the leads, tunneling processes
involving a state with spin $s_{j}=+1$ are depressed. We assume
that the corresponding rates are renormalized by
$0<\gamma_{\uparrow}<1$. With these assumptions, the bi-modal
behavior can be explained. As shown in Fig.~\ref{fig3}b,c where
the linear conductance is plotted, lines with positive slopes are
strongly suppressed because of the tunneling involving states
inside the core of the dot. The lines with a negative slope
exhibit a pronounced bi-modal behavior in the intensity, which
reflects the alternating spin of the $N+1$-th electron in the FD
spectrum, Fig.~\ref{fig3}a.

\begin{figure}
\includegraphics[scale=1]{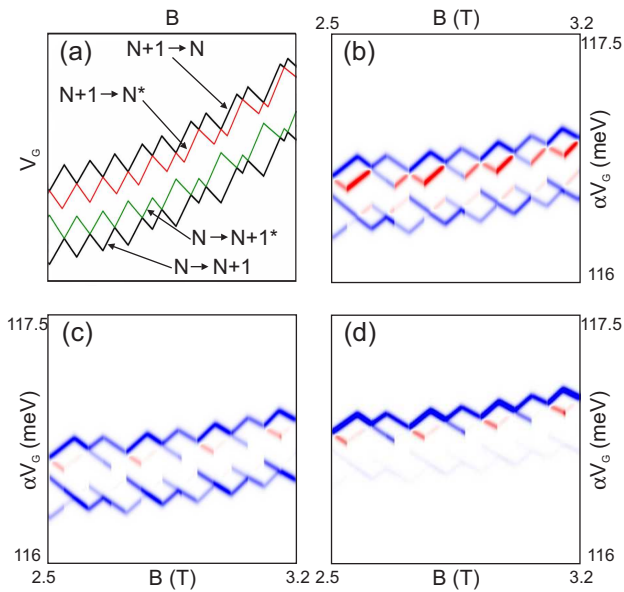}
\caption{(a) Scheme of the nonlinear conductance traces in the
  $(B,V_{\rm G})$ plane in the voltage region of the experiment; (b) nonlinear
  conductance for asymmetric tunnel barriers $\gamma_S/\gamma_D=50$ and
  $\gamma_c^*=\gamma_c$ (red: NDC); (c) nonlinear conductance as in (b) with
  strong relaxation of excited states $r=1.1$ and $\gamma_c^*=0.25\gamma_c$;
  (d) nonlinear conductance as in (c) and asymmetric capacitances
  $C_S/C_D=1/9$.}
\label{fig4}
\end{figure}

At first glance, it might seem that the above assumptions concerning the
tunneling rates are quite exhaustive, and should be sufficient to understand
also the nonlinear behavior at least qualitatively. However, this is not the
case, as we have verified by solving the master equation including the states
relevant to the nonlinear regime. These are the ground states for $N$ and
$N+1$ electrons as in linear transport and their lowest-lying excited
counterparts $N^{*}$ and $N+1^{*}$. For $V>0$, due to the difference in the
chemical potentials of source and drain, the $N\leftrightarrow N+1$
conductance trace splits, giving rise to "stripes" in the $(B,V_{\rm
  G})$-plane (Fig.~\ref{fig4}a). In addition, new traces due to the
transitions $N\leftrightarrow N+1^{*}$ and $N^{*}\leftrightarrow N+1$ appear.
The tunneling rates involving an {\em excited} state of the dot with low
angular momentum are renormalized by the factor $0<\gamma_c^*<1$.

To obtain NDC in the upper part of the stripes (Fig.~\ref{fig2}d, transitions
$N+1\to N$ and $N+1\to N^{*}$) a tunneling asymmetry $\gamma_S/\gamma_D>1$
must be assumed, so that the state $N^{*}$ is dynamically trapped in the dot
\cite{Braggio-PRL-01,Cavaliere-PRL-04,Cavaliere-PRB-04} and the current
decreases. However, this yields as a function of $B$ traces of conductance
peaks with NDC which have both, positive as well as negative slopes
(Fig.~\ref{fig4}b). In contrast to the experimental findings, the NDC features
with positive slopes are even more pronounced than the ones with negative
slope. The tunneling suppression due to dot states with low angular momentum
enhances the dynamical trapping thus favoring NDC. Moreover, in experiment the
transition lines corresponding to transitions $N\to N+1$ and $N\to N+1^{*}$
are clearly suppressed.

For a description consistent with the experiment, parameters have to be chosen
such that the NDC traces with positive slopes vanish. In addition, the
intensity of the lower transition lines must vanish. The former can be
achieved assuming $\gamma_c^*<\gamma_c$, and by assuming a strong relaxation
of the excited states towards the corresponding ground states. We have
introduced a phenomenological relaxation rate $r\Gamma_{0}$ ($r$ dimensionless
constant) into the master equation and consider $r\gg\gamma_c^*$. Then, the
relaxation time is much faster than the dwell time of the low angular momentum
excited states, leading to a substantial reduction of the NDC. The
differential conductance is shown in Fig.~\ref{fig4}c. Due to the strong
relaxation the conductance exhibits spin-driven bi-modal negative-positive
differential conductances in traces with negative slopes, and no NDC in traces
with positive slopes.

In order to obtain vanishing conductance peaks in the lower
stripes, it is necessary to consider asymmetric capacitances. In
Fig.~\ref{fig4}d the differential conductance for the same set of
parameters as in Fig.~\ref{fig4}c is shown, with an asymmetry
$C_S/C_D=1/9$. Qualitative agreement with the experimental
findings is obtained.

We have attempted to reproduce theoretically the experimental
findings by using many different, physically reasonable,
combinations of parameters. However, only with the set of
parameters discussed here we were successful. Apparently, the
patterns of the nonlinear differential conductance traces with the
properties shown in Fig.~\ref{fig2} depend very sensitively on the
properties of the quantum dot and its environment, in addition to
the properties of the tunnel contacts. This has been confirmed
experimentally by measuring the nonlinear differential conductance
of a different 2D quantum dot. In this case, NDC could not be
achieved, although the above discussed bi-modal features of the
linear conductance traces were detected.

In conclusion, we have compared measurements of the nonlinear
conductance spectra of 2D quantum dots in a strong perpendicular
magnetic field with theoretical current-voltage characteristics
obtained by solving a master equation using the Fock-Darwin model
together with a set of parameters that characterize the coupling
between the quantum dot and the leads, and the properties of the
electron states. We have found that in order to reproduce the most
prominent features observed experimentally, a rather complex set
of model parameters is needed. We have found evidence that it is
{\em not} sufficient to assume spin dependent tunneling.
Especially for explaining the regular bi-modal patterns of NDC in
the nonlinear conductance it appears to be necessary to assume
{\em fast relaxation of certain excited quantum dot
  states} in addition to {\em asymmetry of the tunnel rates}. Furthermore, for
reproducing the experimentally observed striking suppression of the traces
corresponding to transitions $N\to (N+1)$ and $N \to (N+1)^{*}$, asymmetry in
source and drain capacitances are necessary.

In contrast to the Coulomb blockade effect, which is universal and
independent of the details of the model in the region of
sequential tunneling, the spin blockade effect appears not to be
universal but very delicately depending on details of the
experimental setup. In addition, the experimental data shown in
Fig.~\ref{fig2} exhibit features, like the behavior of the slopes
of the conductance traces with magnetic field, that apparently
cannot be understood within the Fock-Darwin model including the
spin. For understanding these features, it appears necessary to
take into account electron correlations beyond the constant
interaction model.

Financial support by the German BMBF, the European Union via EU-networks,
HPRN-CT2000-0144, MCRTN-CT2003-504574, and from the Italian MURST PRIN02 is
gratefully acknowledged.

\end{document}